\documentclass{aa}
\usepackage[varg]{txfonts}
\usepackage{graphicx}
\usepackage{natbib,twoopt}
\usepackage[breaklinks=true]{hyperref} %% to avoid \citeads line fills
\usepackage{lscape}
\bibpunct{(}{)}{;}{a}{}{,} %% natbib format for A&A and ApJ
\makeatletter
 \newcommandtwoopt{\citeads}[3][][]{\href{https://ui.adsabs.harvard.edu/abs/#3/abstract}%
 {\def\hyper@linkstart##1##2{}%
 \let\hyper@linkend\@empty\citealp[#1][#2]{#3}}}
 \newcommandtwoopt{\citepads}[3][][]{\href{https://ui.adsabs.harvard.edu/abs/#3/abstract}%
 {\def\hyper@linkstart##1##2{}%
 \let\hyper@linkend\@empty\citep[#1][#2]{#3}}}
 \newcommandtwoopt{\citetads}[3][][]{\href{https://ui.adsabs.harvard.edu/abs/#3/abstract}%
 {\def\hyper@linkstart##1##2{}%
 \let\hyper@linkend\@empty\citet[#1][#2]{#3}}}
 \newcommandtwoopt{\citeyearads}[3][][]%
 {\href{https://ui.adsabs.harvard.edu/abs/#3/abstract}
 {\def\hyper@linkstart##1##2{}%
 \let\hyper@linkend\@empty\citeyear[#1][#2]{#3}}}
\makeatother
\begin{document}

   \title{Dynamics of 2023~FW$_{14}$, the second L$_{4}$ Mars trojan, and a physical 
          characterization using the 10.4~m Gran Telescopio Canarias\thanks{Based on 
          observations made with the Gran Telescopio Canarias (GTC) telescope, in 
          the Spanish Observatorio del Roque de los Muchachos of the Instituto de 
          Astrof\'{\i}sica de Canarias (program ID GTC31-23A).}}
   \author{R.~de~la~Fuente Marcos\inst{1}
            \and
           J.~de~Le\'on\inst{2,3}
            \and 
           C.~de~la~Fuente Marcos\inst{4}
            \and
           M.~R. Alarcon\inst{2,3}
            \and
           J. Licandro\inst{2,3}
            \and
           M. Serra-Ricart\inst{2,3,5}
            \and
           S. Geier\inst{6,2}
            \and
           A. Cabrera-Lavers\inst{6,2,3}
          }
   \authorrunning{R. de la Fuente Marcos et al.}
   \titlerunning{Physical and dynamical characterization of 2023~FW$_{14}$} % Maximum 60 characters
   \offprints{R. de la Fuente Marcos, \email{rauldelafuentemarcos@ucm.es}}
   \institute{AEGORA Research Group,
              Facultad de Ciencias Matem\'aticas,
              Universidad Complutense de Madrid,
              Ciudad Universitaria, E-28040 Madrid, Spain
              \and
              Instituto de Astrof\'{\i}sica de Canarias (IAC),
              C/ V\'{\i}a L\'actea s/n, E-38205 La Laguna, Tenerife, Spain
              \and
              Departamento de Astrof\'{\i}sica, Universidad de La Laguna,
              E-38206 La Laguna, Tenerife, Spain
              \and
              Universidad Complutense de Madrid,
              Ciudad Universitaria, E-28040 Madrid, Spain
              \and
              Light Bridges S.L.,
              Avda. Alcalde Ram\'{\i}rez Bethencourt, 17, E-35004, 
              Las Palmas de Gran Canaria, Canarias, Spain
              \and
              GRANTECAN,
              Cuesta de San Jos\'e s/n, E-38712 Bre\~na Baja, La Palma, Spain
             }
   \date{Received 21 February 2024 / Accepted 4 March 2024}
% Abstract: 300 words
% Text: 3000 words
   \abstract
      {Known Mars trojans could be primordial small bodies that have remained 
       in their present-day orbits for the age of the Solar System. Their 
       orbital distribution is strongly asymmetric; there are over a dozen 
       objects at the L$_{5}$ point and just one at L$_{4}$, (121514) 
       1999~UJ$_{7}$. Most L$_{5}$ trojans appear to form a collision-induced 
       asteroid cluster, known as the Eureka family. Asteroid 2023~FW$_{14}$ 
       was recently discovered and it has a robust orbit determination that 
       may be consistent with a Mars trojan status.
       }
      {Our aim is determine the nature and dynamical properties of 
       2023~FW$_{14}$.
       }
      {We carried out an observational study of 2023~FW$_{14}$ to derive its 
       spectral class using the OSIRIS camera spectrograph at the 10.4~m Gran 
       Telescopio Canarias. We investigated its possible trojan resonance 
       with Mars using direct $N$-body simulations.
       }
      {The reflectance spectrum of 2023~FW$_{14}$ is not compatible with the 
       olivine-rich composition of the Eureka family; it also does not 
       resemble the composition of the Moon, although (101429) 1998~VF$_{31}$ 
       does. The Eureka family and 101429 are at the L$_{5}$ point. The 
       spectrum of 2023~FW$_{14}$ is also different from two out of the three 
       spectra in the literature of the other known L$_{4}$ trojan, 121514, 
       which are of C-type. The visible spectrum of 2023~FW$_{14}$ is 
       consistent with that of an X-type asteroid, as is the third spectrum 
       of 121514. Our calculations confirm that 2023~FW$_{14}$ is the second 
       known L$_{4}$ Mars trojan although it is unlikely to be primordial; it 
       may remain in its present-day ``tadpole'' path for several million 
       years before transferring to a Mars-crossing orbit. It might be a 
       fragment of 121514, but a capture scenario seems more likely. 
       }
      {The discovery of 2023~FW$_{14}$ suggests that regular Mars-crossing 
       asteroids can be captured as temporary Mars trojans. 
       }


   \maketitle

   \section{Introduction\label{Intro}}
      In the Solar System, trojans are small bodies that orbit the Sun engaged in a 1:1 mean-motion resonance with a planet, and
      therefore sharing the values of the orbital period of the planet and its semimajor axis, $a$. Jovian trojans are, for the 
      most part, long-term stable and primordial  (see, e.g., \citealt{1997Natur.385...42L,2020MNRAS.495.4085H}); known Earth 
      trojans are captured near-Earth asteroids and have unstable orbits (see, e.g., \citealt{2011Natur.475..481C,
      2021RNAAS...5...29D,2021ApJ...922L..25H,2022NatCo..13..447S,2022ApJ...938....9Y}). Dynamical studies suggest that some known 
      Mars trojans could have remained in their present-day orbits for the age of the Solar System (see, e.g., 
      \citealt{1994AJ....107.1879M,1999ApJ...517L..63T,2000MNRAS.319...63T,2005P&SS...53..617C,2005Icar..175..397S,
      2013MNRAS.432L..31D}). Most L$_{5}$ Mars trojans are believed to form a collision-induced asteroid cluster,  called the 
      Eureka family \citep{2013Icar..224..144C,2017Icar..293..243C}.

      Although the primordial nature of some of the known Mars trojans is still favored, alternative formation scenarios such as 
      having been ejected from Mars due to a giant impact \citep{2017NatAs...1E.179P} or resulting from rotational-fission via the 
      thermal Yarkovsky-O'Keefe-Radzievskii-Paddack (YORP) mechanism (see, e.g., \citealt{2015Icar..252..339C,
      2020Icar..33513370C}) also agree with the available spectroscopy (see, e.g., \citealt{2017MNRAS.466..489B}). On the 
      other hand, calculations by \citet{2012CeMDA.113...23S} suggest that present-day temporary capture of Mars trojans is 
      possible.

      The recently discovered Amor asteroid 2023~FW$_{14}$ \citep{2023MPEC....G...87C} follows an orbit that resembles those of 
      known Mars trojans. Here we use reflectance spectroscopy and $N$-body simulations to determine the true nature of 
      2023~FW$_{14}$. This letter is organized as follows. In Sect.~\ref{Data} we present the data and tools used in our analyses. 
      In Sect.~\ref{Results} we investigate whether 2023~FW$_{14}$ is a present-day Mars trojan and its origin and future 
      dynamical evolution, and we derive its spectral class. In Sect.~\ref{Discussion} we discuss our results. Our conclusions are 
      summarized in Sect.~\ref{Conclusions}.

   \section{Data and tools\label{Data}}
      Object P21Es0a was found at $w$=21.48~mag by the Panoramic Survey Telescope and Rapid Response System (Pan-STARRS, 
      \citealt{2004SPIE.5489...11K,2013PASP..125..357D}). The first reported observations were carried out by J.~Bulger, T.~Lowe, 
      A.~Schultz, and I.~Smith on March 18, 2023, with the 1.8~m Pan-STARRS~2 Ritchey-Chretien telescope at Haleakala; on April 
      15, 2023, it was announced with the provisional designation 2023~FW$_{14}$ \citep{2023MPEC....G...87C}. On April 19, 2023, a 
      set of precoveries was released \citep{2023MPEC....H..105G}, leading to the orbit determination shown in 
      Table~\ref{elements} as retrieved from the Jet Propulsion Laboratory (JPL) Small-Body Database 
      (SBDB)\footnote{\href{https://ssd.jpl.nasa.gov/tools/sbdb\_lookup.html\#/}
      {https://ssd.jpl.nasa.gov/tools/sbdb\_lookup.html\#/}} provided by the Solar System Dynamics Group (SSDG,
      \citealt{2011jsrs.conf...87G,2015IAUGA..2256293G}).\footnote{\href{https://ssd.jpl.nasa.gov/}{https://ssd.jpl.nasa.gov/}}

%
%------------------------------------------------------------------------------------------------------------------------- TABLE I
%------------------------------------------------------------------------------------------------------ Orbital elements 2023 FW14
%
      \begin{table}
       \centering
       \fontsize{8}{12pt}\selectfont
       \tabcolsep 0.14truecm
       \caption{\label{elements}Values of the heliocentric Keplerian orbital elements of 2023~FW$_{14}$ and their associated 
                1$\sigma$ uncertainties.
               }
       \begin{tabular}{lcc}
        \hline
         Orbital parameter                                 &   & value$\pm$1$\sigma$ uncertainty \\
        \hline
         Semimajor axis, $a$ (au)                          & = &   1.523769939$\pm$0.000000010   \\
         Eccentricity, $e$                                 & = &   0.15811207$\pm$0.00000010     \\
         Inclination, $i$ (\degr)                          & = &  13.272714$\pm$0.000008         \\
         Longitude of the ascending node, $\Omega$ (\degr) & = &  21.84773$\pm$0.00002           \\
         Argument of perihelion, $\omega$ (\degr)          & = & 245.29506$\pm$0.00004           \\
         Mean anomaly, $M$ (\degr)                         & = &  26.03608$\pm$0.00003           \\
         Perihelion distance, $q$ (au)                     & = &   1.2828435$\pm$0.0000002       \\
         Aphelion distance, $Q$ (au)                       & = &   1.764696353$\pm$0.000000012   \\
         Absolute magnitude, $H$ (mag)                     & = &  21.6$\pm$0.4                   \\
        \hline
       \end{tabular}
       \tablefoot{The orbit determination of 2023~FW$_{14}$ is referred to epoch JD 2460200.5 (2023-Sep-13.0) TDB (Barycentric 
                  Dynamical Time, J2000.0 ecliptic and equinox), and is based on 47 observations with a data-arc span of 5503~d 
                  (solution date, April 22, 2023, 07:56:46 PDT). Source: JPL SBDB.
                 }
      \end{table}
%
%---------------------------------------------------------------------------------------------------------------------------------
%

      The orbit determination shown in Table~\ref{elements} is based on 47 observations with a data-arc span of 5503~d or 
      15.07~yr, and corresponds to a near-Earth asteroid (NEA) of the Amor dynamical class with moderate eccentricity, $e$=0.158, 
      and inclination, $i$=13.273\degr; however, its $a$ value matches that of Mars at 1.524~au. The values of $a$ and $e$ 
      automatically make 2023~FW$_{14}$ an object of interest regarding a possible resonant engagement with Mars. In terms of 
      semimajor axis, Mars' co-orbital zone goes from $\sim$1.51645~au to $\sim$1.53095~au (see, e.g., 
      \citealt{2005P&SS...53..617C}). Mars co-orbitals are expected to experience resonant behavior, temporary or long-term, if 
      $e<0.2$. Confirmation of a resonant engagement with Mars requires the analysis of results of $N$-body calculations. The 
      orbit determination in Table~\ref{elements} is referred to standard epoch JD 2460200.5 TDB, which is also the origin of time 
      in the integrations presented here.

      Most L$_{5}$ Mars trojans are thought to be part of an asteroid family. Asteroid family members are expected to have a 
      common origin and composition. Surface mineralogy of asteroids is a proxy for their bulk composition; bodies with a common 
      origin define tight clusters in orbital parameter space. The standard method used to study the surface mineralogy of 
      asteroids is reflectance spectroscopy; the past and future orbital evolution of small bodies is explored using $N$-body 
      simulations.  

      $N$-body simulations require orbit determinations and Cartesian state vectors as input data. In addition, and in order to 
      provide reliable results, these calculations must take into account the uncertainties associated with the orbit 
      determination (see, e.g., \citealt{2018MNRAS.473.2939D,2020MNRAS.494.1089D}). For unstable chaotic dynamical 
      evolution, the results  have to be interpreted statistically. The calculations needed to study the possible resonant status 
      with Mars of 2023~FW$_{14}$ were carried out using a direct $N$-body code developed by \citet{2003gnbs.book.....A},  
      publicly available from the web site of the Institute of Astronomy of the University of 
      Cambridge.\footnote{\href{http://www.ast.cam.ac.uk/~sverre/web/pages/nbody.htm} 
      {http://www.ast.cam.ac.uk/~sverre/web/pages/nbody.htm}} This software applies the Hermite numerical integration scheme 
      devised by \citet{1991ApJ...369..200M}. Extensive results from this code were presented in \citet{2012MNRAS.427..728D}. 

      Calculations were performed in an ecliptic coordinate system with the $X$-axis pointing toward the Vernal Equinox and in the 
      ecliptic plane, the $Z$-axis perpendicular to the ecliptic plane and pointing northward, and the $Y$-axis orthogonal to the 
      previous two and defining a right-handed coordinate system. Our physical model included the eight major planets, the Moon, 
      the barycenter of the Pluto-Charon system, and the three largest asteroids. For accurate initial positions and velocities 
      (see Appendix~\ref{Adata}), we used data from the JPL SSDG {\tt Horizons} online Solar System data and ephemeris 
      computation service,\footnote{\href{https://ssd.jpl.nasa.gov/horizons/}{https://ssd.jpl.nasa.gov/horizons/}} which are based 
      on the DE440/441 planetary ephemeris \citep{2021AJ....161..105P}. Most input data were retrieved from JPL SBDB and 
      {\tt Horizons} using tools provided by the {\tt Python} package {\tt Astroquery} \citep{2019AJ....157...98G} and its 
      {\tt HorizonsClass} class.\footnote{\href{https://astroquery.readthedocs.io/en/latest/jplhorizons/jplhorizons.html}
      {https://astroquery.readthedocs.io/en/latest/jplhorizons/jplhorizons.html}}

      The reflectance spectrum of 2023~FW$_{14}$ was interpreted by performing a taxonomical classification with the help of the 
      {\tt Modeling for Asteroids} (M4AST)\footnote{\href{http://spectre.imcce.fr/m4ast/index.php/index/home}
      {http://spectre.imcce.fr/m4ast/index.php/index/home}} online tool \citep{2012A&A...544A.130P}. After applying a 
      curve-fitting procedure to the spectrum, the tool makes a $\chi^2$ comparison to the taxons defined by 
      \cite{2009Icar..202..160D} and provides the one with the lowest $\chi^2$. 
         
   \section{Results\label{Results}}
      In this section we use reflectance spectroscopy and $N$-body simulations to investigate the surface mineralogy of
      2023~FW$_{14}$ and its resonant status, probable origin, and future orbital evolution.

      \subsection{Spectroscopy}
         The visible spectrum of 2023~FW$_{14}$ was obtained on April 18, 2023, 21:12~UTC, using the Optical System for Imaging 
         and Low Resolution Integrated Spectroscopy (OSIRIS) camera spectrograph \citep{2000SPIE.4008..623C,2010ASSP...14...15C} 
         at the 10.4~m Gran Telescopio Canarias (GTC), located at the El Roque de Los Muchachos Observatory (La Palma, Canary 
         Islands). Observations were done under the program GTC31-23A (PI, J.~de~Le{\'o}n). Appendix~\ref{Aspectrum} details 
         the instrumental setup and data reductions.

%
%---------------------------------------------------------------------------------------------------------------------------------
%
      \begin{figure}
        \centering
         \includegraphics[width=\columnwidth]{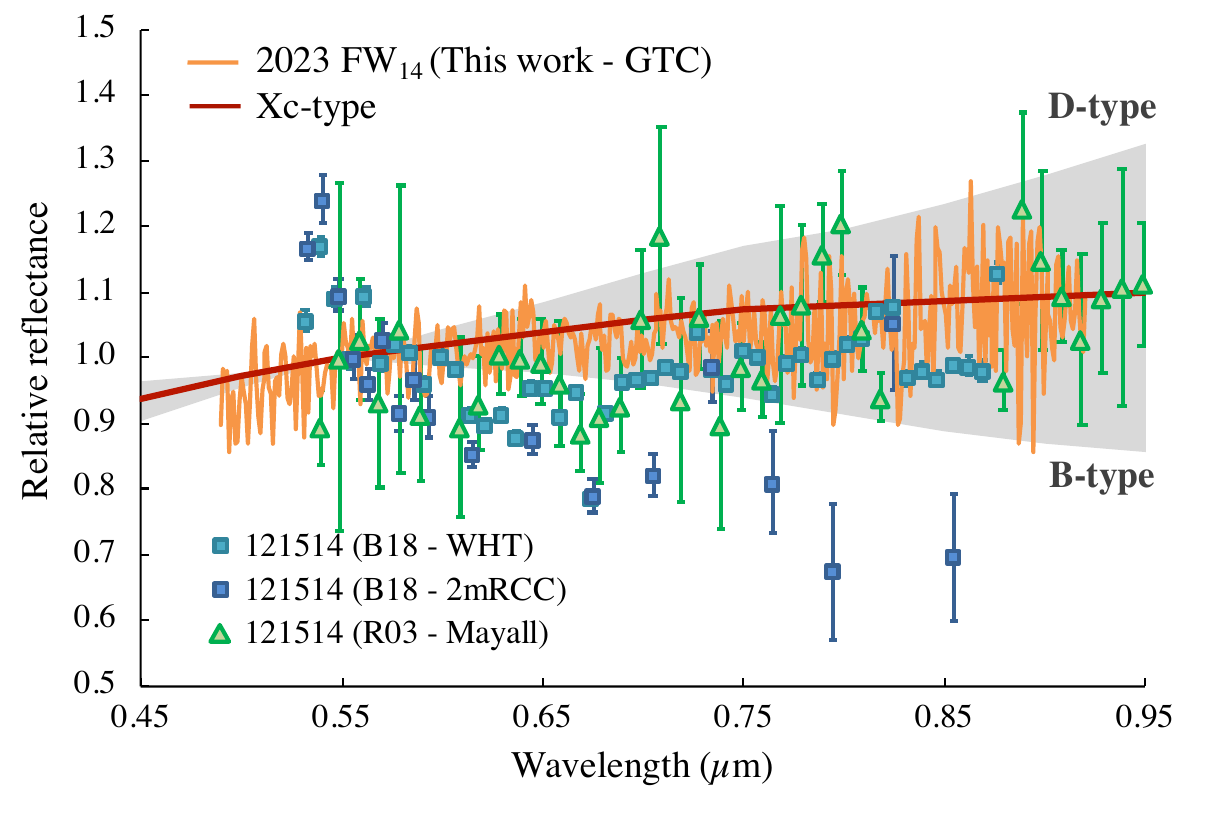}
         \caption{Visible spectrum of 2023~FW$_{14}$ obtained with the 10.4~m GTC (in orange) and its best taxonomical match from 
                  the M4AST online tool, Xc-type (in red). The hatched gray area fills the entire domain between the mean B-type 
                  and D-type classes as defined by \cite{2009Icar..202..160D}. The blue squares correspond to two visible spectra 
                  published in \cite{2018A&A...618A.178B} (labeled B18) of the other known L$_4$ Mars trojan, (121514) 
                  1999~UJ$_{7}$. The third spectrum of 121514, shown here as green triangles, was published in 
                  \cite{2003Icar..165..349R} (labeled R03).}
         \label{spectrum}
      \end{figure}
%
%---------------------------------------------------------------------------------------------------------------------------------
%
         The resulting spectrum is shown in Fig.~\ref{spectrum} (orange). The M4AST online tool provided a taxonomical 
         classification as an Xc-type. For the sake of comparison, we also include the three visible spectra available in the 
         literature of the other known L$_4$ Mars trojan, asteroid (121514) 1999~UJ$_7$: two spectra (in blue) from 
         \cite{2018A&A...618A.178B}, obtained with the 4.2~m William Herschel Telescope (WHT, La Palma, Spain) and the 2~m 
         Ritchey-Chr\'etien-Coud\'e Telescope (2mRCC, Rozhen, Bulgaria), and a third spectrum (in green) from 
         \cite{2003Icar..165..349R}, obtained with the Mayall 4~m telescope (Kitt Peak, Arizona, USA). The third spectrum provides 
         a taxonomical classification as an X-type, while the first spectrum provides a Ch-type classification. According to the 
         authors the spectrum obtained with the 2mRCC telescope had a much poorer quality, and so was not used for classification.

         Additional astrometry and photometric data were obtained with the Two-meter Twin Telescope (TTT), located at the Teide 
         Observatory on the island of Tenerife (Canary Islands, Spain). These are two 0.80~m AltAz telescopes with f/4.4 and f/6.8, 
         respectively. The observations were made using the QHY411M cameras \citep{2023PASP..135e5001A} installed in one of the 
         Nasmyth ports of both telescopes. The data collected served to improve the initial orbit determination. 
         
      \subsection{Resonant status and orbital evolution}
         Trojans appear to move in what are called ``tadpole'' orbits (see, e.g., \citealt{1999ssd..book.....M}) when viewed in a 
         heliocentric frame of reference rotating with the host planet. For values of $e$ and $i$ close to zero, the tadpole 
         trajectory has its center about 60{\degr} ahead of the host planet, around the Lagrange point L$_{4}$ (L$_{4}$ trojan), 
         or follow 60{\degr} behind, around L$_{5}$ (L$_{5}$ trojan). In a general case, when the values of $e$ and/or $i$ are 
         significant, the tadpole center may deviate from the standard +60{\degr} or $-$60{\degr} (or 300{\degr}) locations (see, 
         e.g., \citealt{2000CeMDA..76..131N}). In order to identify trojan resonant behavior, it is necessary to study the 
         evolution of the relative mean longitude $\lambda_{\rm r}=\lambda-\lambda_{\rm P}$, where $\lambda$ and $\lambda_{\rm P}$ 
         are the mean longitudes of the trojan and the host planet, respectively; the relative mean longitude is given by 
         $\lambda=M+\Omega+\omega$, where $M$ is the mean anomaly, $\Omega$ is the longitude of the ascending node, and $\omega$ 
         is the argument of perihelion (see, e.g., \citealt{1999ssd..book.....M}). Only when the value of the critical angle, 
         $\lambda_{\rm r}$, oscillates or librates about +60{\degr} or $-$60{\degr} over an extended period of time, can the small 
         body   be classified as a trojan.

\begin{figure}
        \centering
         \includegraphics[width=\linewidth]{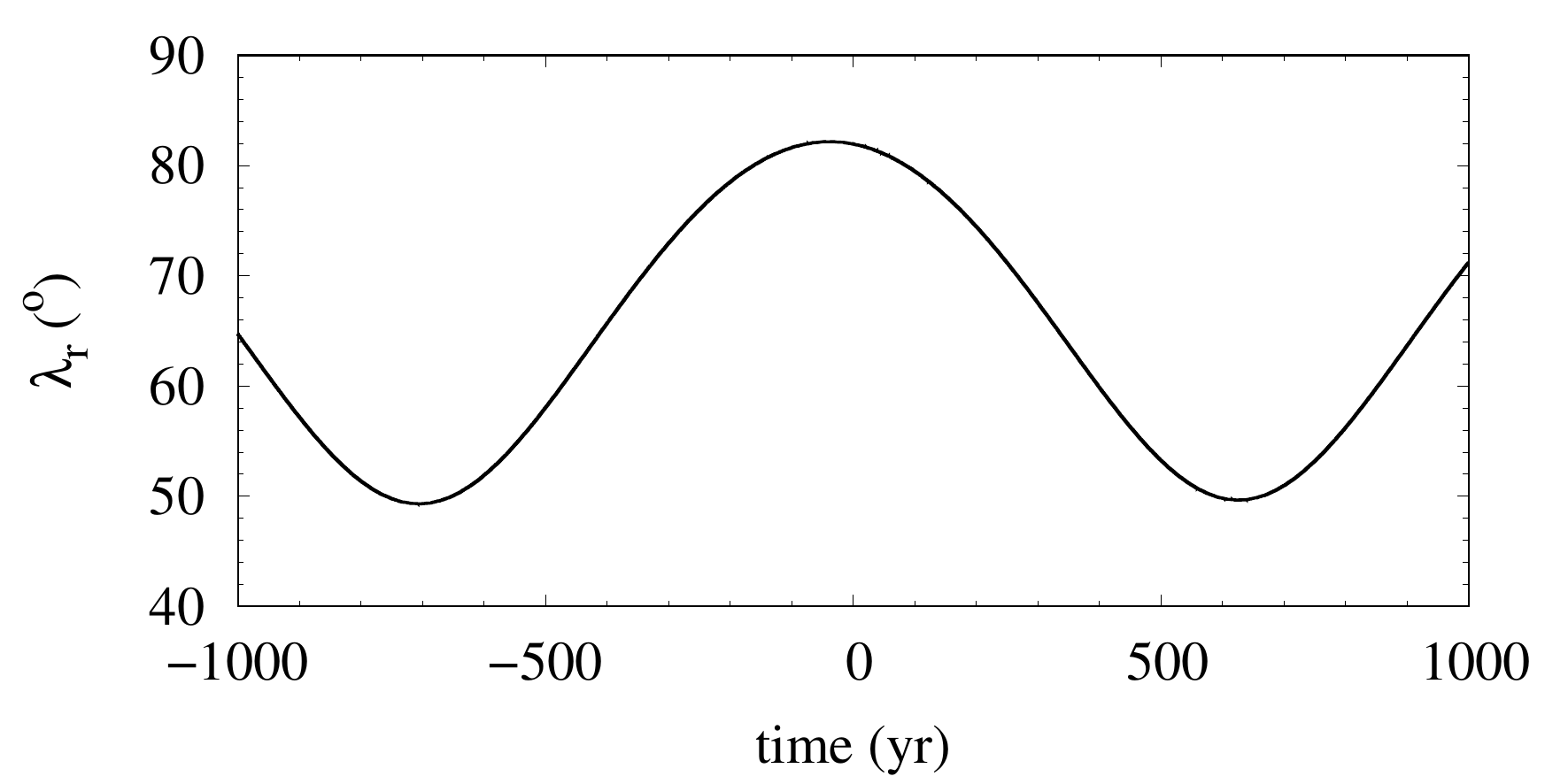}
         \includegraphics[width=\linewidth]{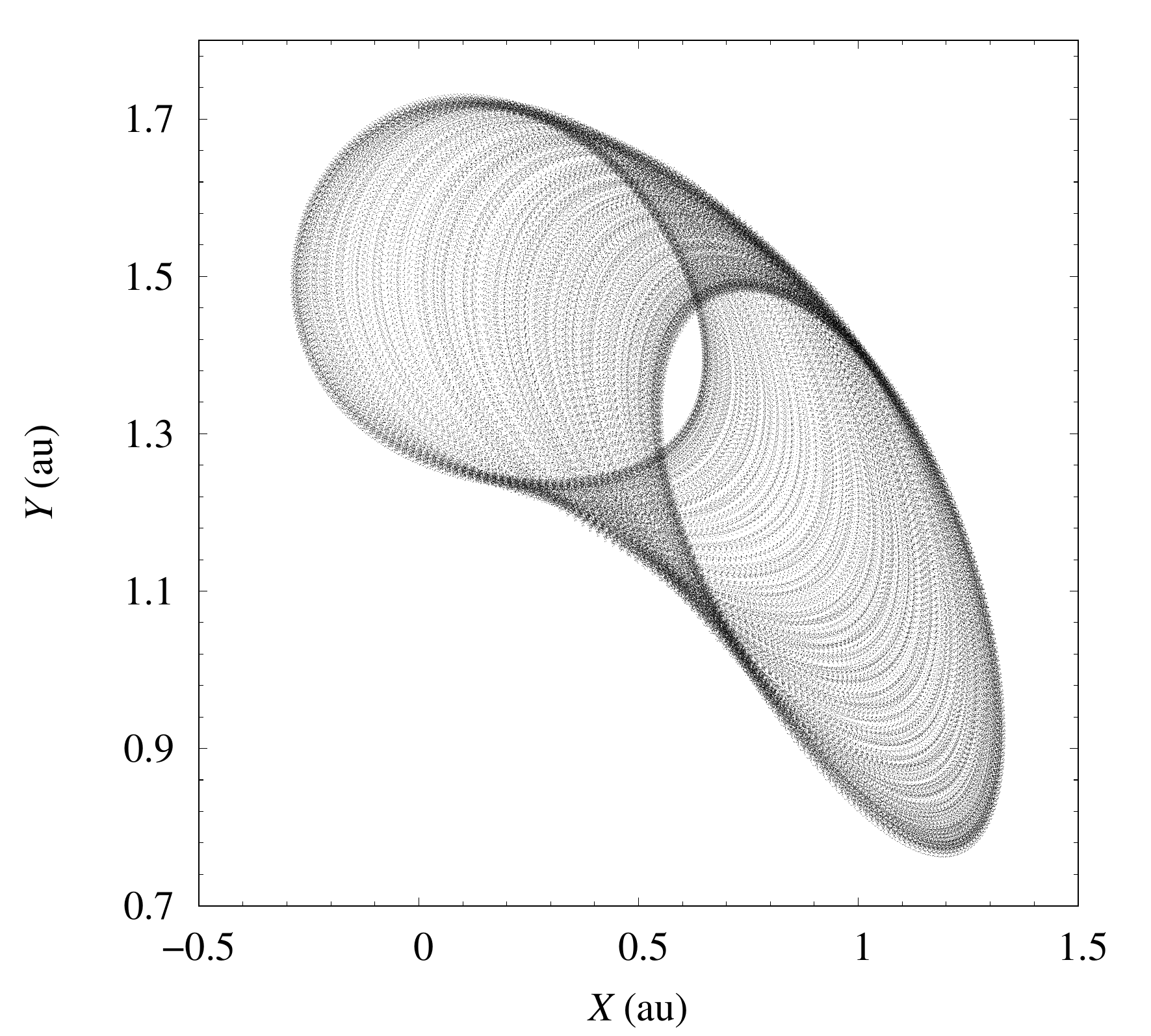}
         \caption{Resonant behavior of 2023~FW$_{14}$. {\it Top panel:} Evolution of the relative mean longitude for the nominal
                  orbit of 2023~FW$_{14}$ in the time interval ($-$1000,~1000)~yr. {\it Bottom panel:} Tadpole loop in the 
                  coordinate system corotating with Mars (Sun-Mars rotating frame) corresponding to the same time interval; 
                  tadpole loops are made of multiple overlapping epicyclic loops. The output time-step size is 0.01~yr.
                 }
         \label{nominal}
      \end{figure}

         The orbit determination in Table~\ref{elements} places 2023~FW$_{14}$ inside of the  Mars co-orbital zone. Therefore, it may 
         be   co-orbital with Mars; in other words, the value of $\lambda_{\rm r}$ may be librating instead of circulating in 
         the interval (0, 2$\pi$). Figure~\ref{nominal}, top panel, shows the evolution of $\lambda_{\rm r}$ for the nominal 
         orbit in Table~\ref{elements}; it displays more than one long period of its librational motion. The period of its L$_4$ 
         trojan motion is 1350~yr, which  is shorter than that of the other known L$_4$ trojan, (121514) 1999~UJ$_7$, 1500~yr, but 
         similar to those of most known L$_5$ trojans \citep{2013MNRAS.432L..31D}. Its amplitude is 33{\degr}, which is smaller than 
         that of 121514, 77{\degr}, and (101429) 1998~VF$_{31}$, 45{\degr}, but larger than those of 5261 Eureka, 11{\degr}; 
         (385250) 2001~DH$_{47}$, 11{\degr}; (311999) 2007 NS$_{2}$, 14{\degr}; 2011~SC$_{191}$, 18{\degr}; 2011~SL$_{25}$, 
         18{\degr}; and 2011~UN$_{63}$, 14{\degr} \citep{2013MNRAS.432L..31D}. The value of $\lambda_r$ oscillates around 
         +66{\degr} instead of +60{\degr} because the orbit of 2023~FW$_{14}$ is somewhat eccentric and inclined. 
         Figure~\ref{nominal}, bottom panel, shows the associated tadpole loop in the coordinate system corotating with Mars. The 
         tadpole loop is the result of the superposition of multiple short-period epicyclic loops reflecting the motion of the 
         trojan relative to Mars. Our short-term integrations of the nominal orbit confirm that 2023~FW$_{14}$ moves in the 
         vicinity of the Mars Lagrange point L$_4$. Its motion about the equilateral libration point ahead of Mars is consistent with 
         trojan dynamical behavior. However, the orbit determination in Table~\ref{elements} is affected by uncertainties, and we 
         must show that the evolution of $\lambda_{\rm r}$ of any orbit statistically consistent with the observational data   
         leads to trojan behavior as well. 
%
%---------------------------------------------------------------------------------------------------------------------------------
%
%
%---------------------------------------------------------------------------------------------------------------------------------
%

\begin{figure}
        \centering
         \includegraphics[width=\linewidth]{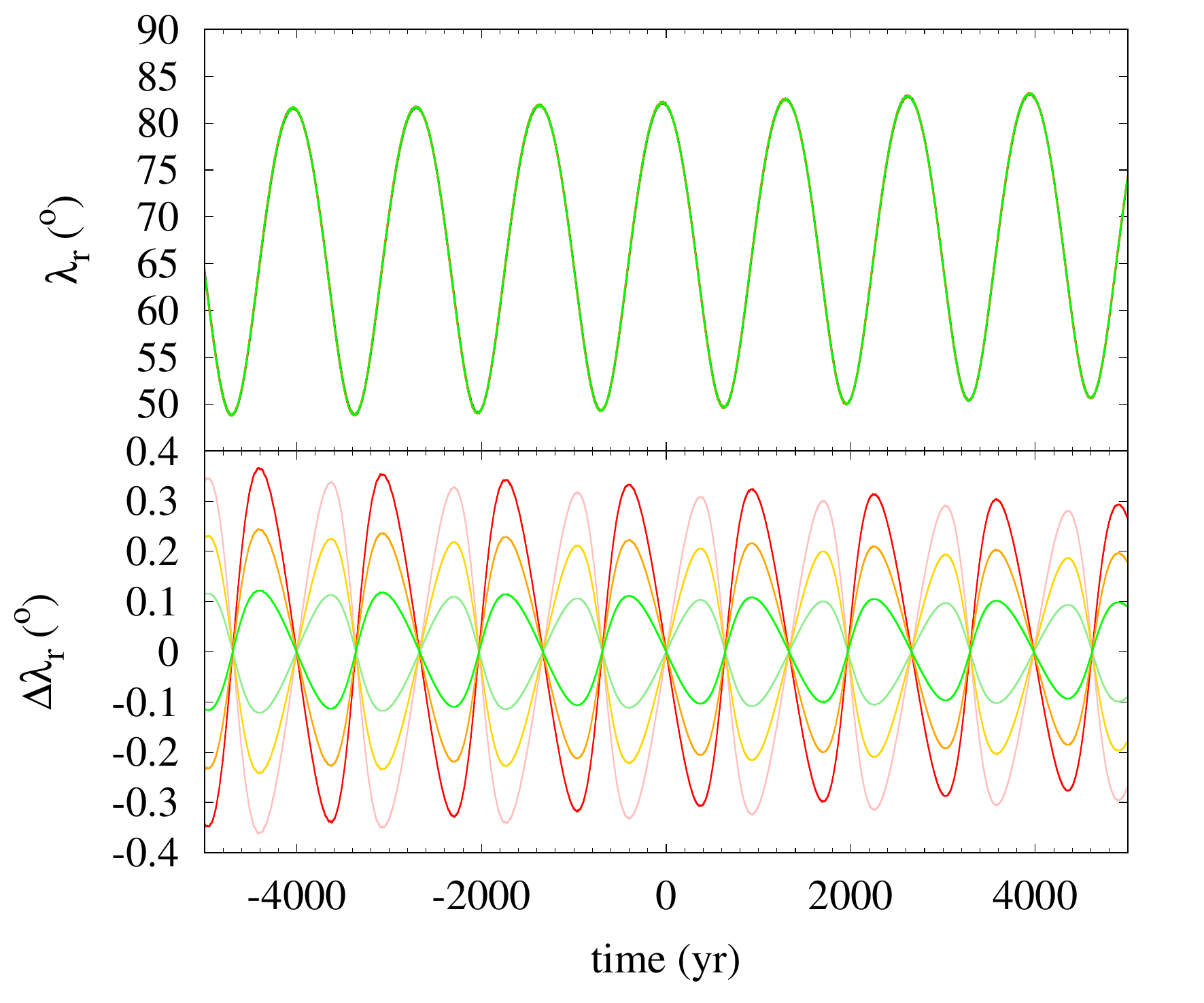}
         \caption{Uncertainties and resonant behavior of 2023~FW$_{14}$. {\it Top panel:} Evolution of the relative mean longitude 
                  with Mars for the nominal orbit (in black) of 2023~FW$_{14}$ and those of relevant control orbits in the time 
                  interval ($-$5\,000,~5\,000)~yr. The control orbits or clones have Cartesian state vectors (see 
                  Appendix~\ref{Adata}) separated $+$3$\sigma$ (green), $-$3$\sigma$ (light-green), $+$6$\sigma$ (orange), 
                  $-$6$\sigma$ (gold), $+$9$\sigma$ (red), and $-$9$\sigma$ (pink) from the nominal values in 
                  Table~\ref{vector2023FW14}. {\it Bottom panel:} Evolution of the difference between the value of the 
                  $\lambda_{\rm r}$ of the control orbits and that of the nominal orbit for the same time interval. The output 
                  time-step size is 0.1~yr.   
                 }
         \label{range}
      \end{figure}

         Figure~\ref{range}, top panel, shows the evolution of $\lambda_{\rm r}$ for the nominal orbit and those of relevant 
         control orbits or clones. The evolution is virtually identical for all the control orbits within $\pm$9$\sigma$ of the 
         nominal orbit determination in Table~\ref{elements}, but the bottom panel shows that when considering the differences in 
         the values of $\lambda_{\rm r}$ with respect to the nominal values, some variation exists. In any case, consistent trojan 
         behavior was found for all the control orbits within $\pm$9$\sigma$ of the nominal orbit determination (1000 were 
         integrated). Therefore, we conclude that 2023~FW$_{14}$ is the second known L$_{4}$ Mars trojan. However, most of the 
         previously known Mars trojans appear to~be stable over the age of the Solar System (see, e.g., 
         \citealt{2013MNRAS.432L..31D}). Longer integrations are needed to investigate whether 2023~FW$_{14}$ is just a temporary 
         trojan or long-term stable.
%
%---------------------------------------------------------------------------------------------------------------------------------
%
%
%---------------------------------------------------------------------------------------------------------------------------------
%

\begin{figure}
        \centering
         \includegraphics[width=\linewidth]{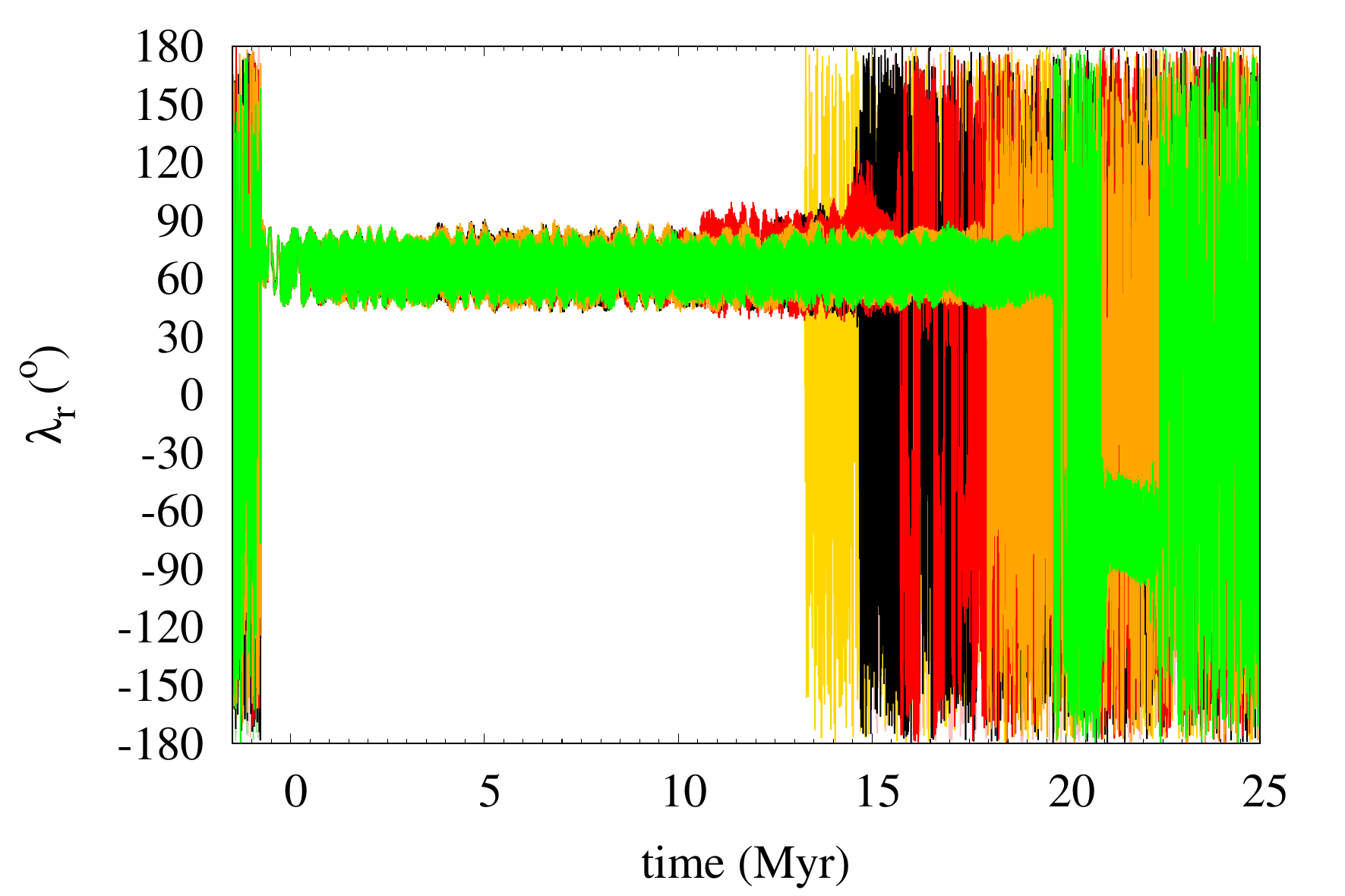}
         \caption{Long-term resonant behavior of 2023~FW$_{14}$. Evolution of the relative mean longitude with Mars for the 
                  nominal orbit (in black) of 2023~FW$_{14}$ and those of relevant control orbits in the time interval 
                  ($-$1.5,~25.0)~Myr. The control orbits or clones have Cartesian state vectors (see Appendix~\ref{Adata}) 
                  separated $+$3$\sigma$ (in green), $-$3$\sigma$ (in light-green), $+$6$\sigma$ (in orange), $-$6$\sigma$ (in 
                  gold), $+$9$\sigma$ (in red), and $-$9$\sigma$ (in pink) from the nominal values in Table~\ref{vector2023FW14}.
                  The output time-step size is 5\,000~yr. 
                 }
         \label{rangelongterm}
      \end{figure}

         Figure~\ref{rangelongterm} shows the evolution of $\lambda_{\rm r}$ for the nominal orbit and those of relevant control 
         orbits for longer integrations. The evolution of $\lambda_{\rm r}$ confirms that 2023~FW$_{14}$ is not a long-term stable 
         L$_{4}$ Mars trojan. It also shows that the evolution of this temporary trojan is far more unstable when considering 
         its past. The object only became an L$_{4}$ Mars trojan nearly 1~Myr ago and it will leave its current trojan engagement 
         with Mars perhaps as early as 13~Myr from now. The most straightforward interpretation of these results is that it might 
         have been captured from the population of Mars-crossing NEAs and will return to it after escaping Mars' co-orbital 
         region. Although all the control orbits became trojans at about the same time when integrated backward in time, there is 
         a considerable dispersion in the results of forward integrations (although the evolutions of 7 orbits are shown, 25 
         were studied). We can conclude with certainty that 2023~FW$_{14}$ is not a long-term stable L$_{4}$ Mars trojan, but we 
         cannot provide a reliable prediction of the exact duration of the current trojan episode beyond stating that it will last 
         more than 10~Myr. Therefore, it is unlikely to be primordial.

%
%---------------------------------------------------------------------------------------------------------------------------------
%
%
%---------------------------------------------------------------------------------------------------------------------------------
%

   \section{Discussion\label{Discussion}}
      Among L$_{4}$ Mars trojans, 2023~FW$_{14}$ has the highest orbital eccentricity (0.158) and the lowest inclination 
      (13.273\degr). This value of $i$ places 2023~FW$_{14}$ inside the unstable region identified by \citet{2005Icar..175..397S} 
      where various secular resonances will remove a trojan within a few million years. In addition, it has the largest value of 
      $H$ (21.6~mag). NEOWISE observations of (121514) 1999~UJ$_{7}$ \citep{2016AJ....152...63N} provide a value of its visible 
      albedo of 0.047$\pm$0.023, which is compatible with its classification as an X-type asteroid by \citet{2003Icar..165..349R}. 
      Using this value of the albedo and the absolute magnitude in Table~\ref{elements}, we derive a mean diameter of 
      $D=318_{-199}^{+493}$~m for 2023~FW$_{14}$. If the only other L$_{4}$ Mars trojan, 121514, is the largest known Mars trojan, 
      then 2023~FW$_{14}$ could be one of the smallest known so far. 

      In principle, the long-term behavior into the past of 2023~FW$_{14}$ is compatible with capture from the population of 
      Mars-crossing NEAs, but an origin as a fragment of another trojan, either known or still undiscovered, cannot be discarded 
      considering the available data. Our calculations indicate that the evolution of 2023~FW$_{14}$ is stable for over 10~Myr; 
      this is much shorter than the stability timescale of the other trojans, but also significantly longer than the typical 
      duration of the resonant episodes of transient Mars co-orbitals discussed by \citet{2005P&SS...53..617C}. This might be 
      hinting at an in situ origin for 2023~FW$_{14}$. Regarding the provenance of 121514, it has been suspected that this trojan 
      is not primordial, but it was captured about 4~Gyr ago \citep{2013MNRAS.432L..31D}. Another probable captured trojan is 
      (101429) 1998~VF$_{31}$ \citep{2013MNRAS.432L..31D}. Spectroscopic results also argue for a different origin in the case of 
      101429 and 121514 \citep{2007Icar..192..434R,2021Icar..35413994C}. 

      The second hypothesis regarding the origin of 2023~FW$_{14}$ can be partially tested using the spectroscopic information. 
      The reflectance spectrum of 2023~FW$_{14}$ is neither compatible with an olivine-rich composition like that of the Eureka 
      family \citep{2017MNRAS.466..489B} nor resembles the one of the Moon, like in the case of 101429 
      \citep{2021Icar..35413994C}, both at L$_{5}$. It is on the contrary a primitive-like spectrum, matching the Xc-type class. 
      There are three spectra in the literature for the other known L$_4$ trojan, asteroid 121514. The spectrum obtained by 
      \cite{2018A&A...618A.178B} with the 4.2~m WHT yields a Ch-type classification, mainly due to the presence of a broad 
      absorption feature centered at $\sim$0.64--0.65~$\mu$m. However, neither the center nor the shape of this absorption is 
      typical of the 0.7~$\mu$m absorption observed in Ch-type asteroids and associated with phyllosilicates. Even the authors 
      note this point and invoke a potential new subclass of the C-complex to explain this spectrum. The other spectrum of 
      121514, classified as an X-type, was obtained by \cite{2003Icar..165..349R}. All in all, and even considering the poorer 
      quality of the data previously obtained for 121514 compared to our spectrum of 2023~FW$_{14}$, we can confidently say that 
      these two asteroids have primitive-like spectra, in contrast with the trojans studied in L$_5$. Although incomplete, the 
      data support the interpretation of 2023~FW$_{14}$ as an interloper captured from the Mars-crossing NEA population, but they
      cannot be used to reject the competing hypothesis that 2023~FW$_{14}$ was produced in situ by, for example, YORP-induced 
      rotational fission, as perhaps in the case of the Eureka family \citep{2015Icar..252..339C,2020Icar..33513370C} because the 
      spectra of 121514 and 2023~FW$_{14}$ are somewhat close (see Fig.~\ref{spectrum}). On the other hand, and although the 
      present-day 121514 is a slow rotator \citep{2018A&A...618A.178B} and therefore not capable of shedding material via the YORP 
      mechanism, its spin state might have been different in the past \citep{2020Icar..33513370C}.

   \section{Summary and conclusions\label{Conclusions}}
      In this letter we presented spectroscopic observations of Mars' second L$_{4}$ trojan, 2023~FW$_{14}$, obtained on April 18, 
      2023, using the OSIRIS camera-spectrograph at the 10.4~m GTC. We used the spectrum to provide a physical characterization of 
      the object and direct $N$-body simulations to confirm its trojan resonant state and investigate its orbital evolution. Our 
      conclusions can be summarized as follows:
      \begin{enumerate}
         \item We find that 2023~FW$_{14}$ has a visible spectrum consistent with that of an Xc-type asteroid. 
         \item We confirm that 2023~FW$_{14}$ is the second known L$_{4}$ Mars trojan.
         \item We confirm that 2023~FW$_{14}$ is a temporary L$_{4}$ Mars trojan that might have been captured from the 
               Mars-crossing NEA population about 1~Myr ago or, less likely, shed from (121514) 1999~UJ$_{7}$. Its current trojan 
               episode will last at least 10~Myr.
      \end{enumerate}
      \citet{2012CeMDA.113...23S} found that the present-day temporary capture of Mars trojans is possible. The discovery of 
      2023~FW$_{14}$ could be the confirmation of this theoretical possibility.

   \begin{acknowledgements}
      We thank the anonymous referee for a prompt and helpful report, and S. Deen for finding precovery images of 2023~FW$_{14}$
      that improved the orbital solution of this object significantly and for additional comments. RdlFM and CdlFM thank S.~J. 
      Aarseth for providing one of the codes used in this research and A.~I. G\'omez de Castro for providing access to computing 
      facilities. JdL and JL acknowledge financial support from the Spanish Ministry of Science and Innovation (MICINN) through 
      the Spanish State Research Agency, under Severo Ochoa Programme 2020-2023 (CEX2019-000920-S). This work was partially 
      supported by the Spanish `Agencia Estatal de Investigaci\'on (Ministerio de Ciencia e Innovaci\'on)' under grant 
      PID2020-116726RB-I00 /AEI/10.13039/501100011033. Based on observations made with the Gran Telescopio Canarias (GTC), 
      installed at the Spanish Observatorio del Roque de los Muchachos of the Instituto de Astrof\'{\i}sica de Canarias, on the 
      island of La Palma. This work is partly based on data obtained with the instrument OSIRIS, built by a Consortium led by the 
      Instituto de Astrof\'{\i}sica de Canarias in collaboration with the Instituto de Astronom\'{\i}a of the Universidad Nacional 
      Aut\'onoma de Mexico. OSIRIS was funded by GRANTECAN and the National Plan of Astronomy and Astrophysics of the Spanish 
      Government. This letter includes observations made with the Two-meter Twin Telescope (TTT) at the IAC's Teide Observatory 
      that Light Bridges, SL, operates on the Island of Tenerife, Canary Islands (Spain). The Observing Time Rights (DTO) used for 
      this research at the TTT have been provided by the Instituto de Astrof\'{\i}sica de Canarias. In preparation of this letter, 
      we made use of the NASA Astrophysics Data System, the ASTRO-PH e-print server, and the MPC data server. 
   \end{acknowledgements}

   \bibliographystyle{aa}

\begin{thebibliography}{}
      \bibitem[Aarseth(2003)]{2003gnbs.book.....A} Aarseth, S.~J. 2003,
              Gravitational N-Body Simulations
              (Cambridge: Cambridge University Press), 27
      \bibitem[Alarcon et al.(2023)]{2023PASP..135e5001A} Alarcon, M.~R., Licandro, J., Serra-Ricart, M., et al.\ 2023,
              \pasp, 135, 055001
      \bibitem[Borisov et al.(2017)]{2017MNRAS.466..489B} Borisov, G., Christou, A., Bagnulo, S., et al.\ 2017, 
              \mnras, 466, 489
      \bibitem[Borisov et al.(2018)]{2018A&A...618A.178B} Borisov, G., Christou, A.~A., Colas, F., et al.\ 2018, 
              \aap, 618, A178
      \bibitem[Cepa et al.(2000)]{2000SPIE.4008..623C} Cepa, J., Aguiar, M., Escalera, V.~G., et al.\ 2000, 
              \procspie, 4008, 623
      \bibitem[Cepa(2010)]{2010ASSP...14...15C} Cepa, J.\ 2010, 
              Astrophysics and Space Science Proceedings, 14, 15
      \bibitem[Chambers et al.(2023)]{2023MPEC....G...87C} Chambers, K., Chastel, S., de Boer, T., et al.\ 2023, 
              Minor Planet Electronic Circulars, 2023-G87
      \bibitem[Christou(2013)]{2013Icar..224..144C} Christou, A.~A.\ 2013, 
              \icarus, 224, 144
      \bibitem[Christou et al.(2017)]{2017Icar..293..243C} Christou, A.~A., Borisov, G., Dell'Oro, A., et al.\ 2017, 
              \icarus, 293, 243
      \bibitem[Christou et al.(2020)]{2020Icar..33513370C} Christou, A.~A., Borisov, G., Dell'Oro, A., et al.\ 2020, 
              \icarus, 335, 113370
      \bibitem[Christou et al.(2021)]{2021Icar..35413994C} Christou, A.~A., Borisov, G., Dell'Oro, A., et al.\ 2021, 
              \icarus, 354, 113994
      \bibitem[Connors et al.(2011)]{2011Natur.475..481C} Connors, M., Wiegert, P., \& Veillet, C.\ 2011, 
              \nat, 475, 481
      \bibitem[Connors et al.(2005)]{2005P&SS...53..617C} Connors, M., Stacey, G., Brasser, R., et al.\ 2005, 
              \planss, 53, 617
      \bibitem[{\'C}uk et al.(2015)]{2015Icar..252..339C} {\'C}uk, M., Christou, A.~A., \& Hamilton, D.~P.\ 2015, 
              \icarus, 252, 339
      \bibitem[de la Fuente Marcos \& de la Fuente Marcos(2012)]{2012MNRAS.427..728D} de la Fuente Marcos, C. \& de la Fuente Marcos, R.\ 2012, 
              \mnras, 427, 728
      \bibitem[de la Fuente Marcos \& de la Fuente Marcos(2013)]{2013MNRAS.432L..31D} de la Fuente Marcos, C. \& de la Fuente Marcos, R.\ 2013, 
              \mnras, 432, L31
      \bibitem[de la Fuente Marcos \& de la Fuente Marcos(2018)]{2018MNRAS.473.2939D} de la Fuente Marcos, C. \& de la Fuente Marcos, R.\ 2018, 
              \mnras, 473, 2939
      \bibitem[de la Fuente Marcos \& de la Fuente Marcos(2020)]{2020MNRAS.494.1089D} de la Fuente Marcos, C. \& de la Fuente Marcos, R.\ 2020, 
              \mnras, 494, 1089 
      \bibitem[de la Fuente Marcos \& de la Fuente Marcos(2021)]{2021RNAAS...5...29D} de la Fuente Marcos, C. \& de la Fuente Marcos, R.\ 2021, 
              RNAAS, 5, 29
      \bibitem[DeMeo et al.(2009)]{2009Icar..202..160D} DeMeo, F., Binzel, R. P., Slivan, S. M., et al.\ 2009, 
              \icarus, 202, 160
      \bibitem[Denneau et al.(2013)]{2013PASP..125..357D} Denneau, L., Jedicke, R., Grav, T., et al.\ 2013, 
              \pasp, 125, 357
      \bibitem[Ginsburg et al.(2019)]{2019AJ....157...98G} Ginsburg, A., Sip{\H{o}}cz, B.~M., Brasseur, C.~E., et al.\ 2019, 
              \aj, 157, 98
      \bibitem[{{Giorgini}(2011)}]{2011jsrs.conf...87G} {Giorgini}, J. 2011,
              in Journ\'ees Syst\`emes de R\'ef\'erence Spatio-temporels 2010,
              ed. N.~{Capitaine}, 87--87
      \bibitem[Giorgini(2015)]{2015IAUGA..2256293G} Giorgini, J.~D.\ 2015,
              IAUGA, 22, 2256293
      \bibitem[Gray et al.(2023)]{2023MPEC....H..105G} Gray, B., Rankin, D., Shelly, F.~C., et al.\ 2023, 
              Minor Planet Electronic Circulars, 2023-H105
      \bibitem[Holt et al.(2020)]{2020MNRAS.495.4085H} Holt, T.~R., Nesvorn{\'y}, D., Horner, J., et al.\ 2020, 
              \mnras, 495, 4085
      \bibitem[Hui et al.(2021)]{2021ApJ...922L..25H} Hui, M.-T., Wiegert, P.~A., Tholen, D.~J., et al.\ 2021, 
              \apjl, 922, L25
      \bibitem[Kaiser(2004)]{2004SPIE.5489...11K} Kaiser, N.\ 2004, 
              \procspie, 5489, 11
      \bibitem[Levison et al.(1997)]{1997Natur.385...42L} Levison, H.~F., Shoemaker, E.~M., \& Shoemaker, C.~S.\ 1997, 
              \nat, 385, 42
      \bibitem[Licandro et al.(2019)]{2019A&A...625A.133L} Licandro, J., de la Fuente Marcos, C., de la Fuente Marcos, R., et al.\ 2019, 
              \aap, 625, A133
      \bibitem[Makino(1991)]{1991ApJ...369..200M} Makino, J.\ 1991,
              \apj, 369, 200
      \bibitem[Mikkola \& Innanen(1994)]{1994AJ....107.1879M} Mikkola, S. \& Innanen, K.\ 1994, 
              \aj, 107, 1879
      \bibitem[Murray \& Dermott(1999)]{1999ssd..book.....M} Murray, C.~D., \& Dermott, S.~F.\ 1999,
              Solar System Dynamics
              (Cambridge: Cambridge University Press)
      \bibitem[Namouni \& Murray(2000)]{2000CeMDA..76..131N} Namouni, F. \& Murray, C.~D.\ 2000, 
              Celestial Mechanics and Dynamical Astronomy, 76, 131
      \bibitem[Nugent et al.(2016)]{2016AJ....152...63N} Nugent, C.~R., Mainzer, A., Bauer, J., et al.\ 2016, 
              \aj, 152, 63
      \bibitem[Park et al.(2021)]{2021AJ....161..105P} Park, R.~S., Folkner, W.~M., Williams, J.~G., et al.\ 2021, 
              \aj, 161, 105 
      \bibitem[Polishook et al.(2017)]{2017NatAs...1E.179P} Polishook, D., Jacobson, S.~A., Morbidelli, A., et al.\ 2017, 
              Nature Astronomy, 1, 0179
      \bibitem[Popescu et al.(2012)]{2012A&A...544A.130P} Popescu, M., Birlan, M., \& Nedelcu, D.~A.\ 2012, 
              \aap, 544, A130
      \bibitem[Rivkin et al.(2003)]{2003Icar..165..349R} Rivkin, A. S., Binzel, R. P., Howell, E. S., et al.\ 2003,
              \icarus, 165, 349
      \bibitem[Rivkin et al.(2007)]{2007Icar..192..434R} Rivkin, A.~S., Trilling, D.~E., Thomas, C.~A., et al.\ 2007, 
              \icarus, 192, 434
      \bibitem[Santana-Ros et al.(2022)]{2022NatCo..13..447S} Santana-Ros, T., Micheli, M., Faggioli, L., et al.\ 2022, 
              Nat. Commun., 13, 447
      \bibitem[Scholl et al.(2005)]{2005Icar..175..397S} Scholl, H., Marzari, F., \& Tricarico, P.\ 2005, 
              \icarus, 175, 397
      \bibitem[Schwarz \& Dvorak(2012)]{2012CeMDA.113...23S} Schwarz, R. \& Dvorak, R.\ 2012, 
              Celestial Mechanics and Dynamical Astronomy, 113, 23
      \bibitem[Tabachnik \& Evans(1999)]{1999ApJ...517L..63T} Tabachnik, S. \& Evans, N.~W.\ 1999, 
              \apjl, 517, L63
      \bibitem[Tabachnik \& Evans(2000)]{2000MNRAS.319...63T} Tabachnik, S.~A. \& Evans, N.~W.\ 2000, 
              \mnras, 319, 63
      \bibitem[Yeager \& Golovich(2022)]{2022ApJ...938....9Y} Yeager, T. \& Golovich, N.\ 2022, 
              \apj, 938, 9
   \end{thebibliography}

   \begin{appendix}
      \section{Spectroscopic observations and data reduction\label{Aspectrum}}
         We used the OSIRIS camera-spectrograph at the 10.4~m GTC. The OSIRIS detector is a blue-sensitive monolithic 
         4096$\times$4096 pixel CCD that provides an unvigneted field of view of 7.8'$\times$7.8'. The standard operation mode of 
         the instrument uses a 2$\times$2 binning. We used the R300R grism that covers a wavelength range from 0.48 to 
         0.92~$\mu$m, with a dispersion of 7.74~\AA/pixel for a 0.6" slit. We used the 1.2" slit, oriented to the parallactic 
         angle, and with the tracking of the telescope at a set rate matching the proper motion of the asteroid. We obtained three 
         consecutive spectra of 900~s of exposure time each, at an airmass of 1.3, offsetting the telescope 10" in the slit 
         direction between the spectra. To obtain the reflectance spectra of the asteroid, we also observed two solar analog stars 
         (Landolt SA 98-978 and SA 102-1081), using the same instrumental configuration as for the asteroid, and at a similar 
         airmass. In the case of the stars, we obtained three individual spectra, also offsetting the telescope in the slit 
         direction by 10" between individual spectra. Spectral images of the asteroid and the solar analog stars were bias and 
         flat-field corrected. The 2D spectra were background subtracted and collapsed to 1D by adding all the flux within an 
         aperture (typically defined as the distance from the center of the spatial profile where the intensity is 10\% of the 
         peak intensity). One-dimensional spectra where then wavelength calibrated using Xe+Ne+HgAr arc lamps. We added the three 
         asteroid spectra, and averaged, for each solar analog, the corresponding individual spectra. Then, as a final step, we 
         divided the spectrum of the asteroid by the spectrum of each solar analog star, and averaged the two resulting ratios. 
         This  final spectrum is  shown in Fig.~\ref{spectrum} in orange. Additional details are described  in 
         \cite{2019A&A...625A.133L}, among others.

      \section{Input data\label{Adata}}
         Here, we include the barycentric Cartesian state vector of L$_{4}$ Mars trojan 2023~FW$_{14}$. This vector and its 
         uncertainties have been used to perform the calculations discussed in the sections and to generate the figures that 
         display the time evolution of the critical angle, $\lambda_{\rm r}$. For example, a new value of the $X$-component of the 
         state vector is computed as $X_{\rm c} = X + \sigma_X \ r$, where $r$ is an univariate Gaussian random number, and $X$ 
         and $\sigma_X$ are the mean value and its 1$\sigma$ uncertainty from  Table~\ref{vector2023FW14}.
%
%------------------------------------------------------------------------------------------------------------------------- TABLE II
%--------------------------------------------------------------------------------------- Geometric Cartesian state vector 2023 FW14
%
     \begin{table}
      \centering
      \fontsize{8}{12pt}\selectfont
      \tabcolsep 0.15truecm
      \caption{\label{vector2023FW14}Barycentric Cartesian state vector of 2023~FW$_{14}$: components and associated 1$\sigma$ 
               uncertainties.
              }
      \begin{tabular}{ccc}
       \hline
        Component                         &   &    value$\pm$1$\sigma$ uncertainty                                \\
       \hline
        $X$ (au)                          & = &    6.915652422725076$\times10^{-1}$$\pm$4.45004793$\times10^{-7}$ \\
        $Y$ (au)                          & = & $-$1.076839004164107$\times10^{+0}$$\pm$3.69180966$\times10^{-7}$ \\
        $Z$ (au)                          & = & $-$2.965493352209987$\times10^{-1}$$\pm$1.79345340$\times10^{-7}$ \\
        $V_X$ (au/d)                      & = &    1.411478671143701$\times10^{-2}$$\pm$1.85662865$\times10^{-9}$ \\
        $V_Y$ (au/d)                      & = &    7.479403282969601$\times10^{-3}$$\pm$4.72439409$\times10^{-9}$ \\
        $V_Z$ (au/d)                      & = &    4.005959201684231$\times10^{-4}$$\pm$1.69044868$\times10^{-9}$ \\
       \hline
      \end{tabular}
      \tablefoot{Data are referred to epoch JD 2460200.5, which corresponds to 0:00 on 2023-Sep-13.0 TDB (J2000.0 ecliptic and 
                 equinox). Source: JPL {\tt Horizons}.
                }
     \end{table}
%
%---------------------------------------------------------------------------------------------------------------------------------
%

   \end{appendix}

\end{document}